\documentclass[12pt]{article}                    
\usepackage{graphicx}
\usepackage{mathptmx}      
\usepackage{amssymb}
\usepackage{amsmath}
\usepackage{amsmath,amsbsy,amsfonts,amssymb,graphics,epsfig,float,bm,
verbatim,subfigure}

\newcommand{\la}{\label}
\newcommand{\be}{\begin{equation}}
\newcommand{\en}{\end{equation}}

\renewcommand{\vec}[1]{\boldsymbol{#1}}

\newcommand{\ee}{\textrm{e}}

\begin{document}

\title{On the Abaqus FEA model of finite viscoelasticity}

\author{Jacopo Ciambella$^a$, Michel Destrade$^b$, Ray W. Ogden$^c$\\[12pt]
$^a$Department of Structural and Geotechnical Engineering, \\
Universit\`a di Roma ``La Sapienza'',\\
Via Eudossiana, 18 - 00184 - Roma, Italy.\\[12pt]
$^b$              Institut Jean Le Rond d'Alembert, \\
CNRS/Universit\'e Pierre et Marie Curie, \\
Case 162, 75252 Paris Cedex 05, France.\\[12pt]
$^c$Department of Mathematics, \\University of Glasgow,\\
Glasgow G128QW, United Kingdom.}

\date{}

\maketitle

\bigskip


\begin{abstract}

Predictions of the QLV (Quasi-Linear Viscoelastic) constitutive law are compared with those of the ABAQUS viscoelastic
model for two simple motions in order to highlight, in particular, their very different dissipation rates and certain
shortcomings of the ABAQUS model.

\end{abstract}
\newpage


\section{Introduction}
\label{intro}


As the demand for fuel efficient cars intensifies, tyre manufacturers
try harder to estimate accurately 
the energy losses of their products. 
However the mechanics of tires are a most complex topic due to the
variety of mechanical factors 
involved, such as nonlinear responses, structural inhomogeneities,
complicated geometries and 
solicitations, hypo-, visco-, and hyperelasticity, etc. 
Consequently, advanced finite element codes are often called to the
rescue in order to simulate 
real-world situations, and to evaluate the energy efficiency of a tyre.
These commercial codes are often used as a ``black box'', and the
validity of the answer is rarely put into 
question, even though it might provide a decisive argument in favor of,
or against, the viability of a given 
tyre model. 

Here we examine the current implementation of nonlinear viscoelastic
effects in the Abaqus Finite Element Analysis (FEA) package. 
We find that it is not sound physically, and that it gives results that
are not consistent with the 
thermomechanically-based Quasi Linear Viscoeleastic (QLV) model. 
We present both models in the next section, and highlight their main
differences. 
Then we investigate two toy experiments, using the incompressible
neo-Hookean material for the elastic 
response. 
We consider the equi-biaxial extension test in Section 3, and the simple
shear test in Section 4. 
In both cases, the Abaqus FEA model predicts higher energy losses than
the QLV model.


\section{Abaqus FEA finite viscoelasticity model}


First we introduce some notations. 
We call $\vec{F}$ the deformation gradient;
it is defined as $\vec{F} = \partial \vec{x}/\vec{X}$, where $\vec{x}$
is the coordinate in the current 
configuration of a particle at $\vec{X}$ in the reference
configuration. 
We call $J$ its determinant: $J = \text{det}\vec{F}$, and 
$\vec{B} \equiv \vec{F F}^T$,
$\vec{C} \equiv \vec{F}^T\vec{F}$ the associated Cauchy-Green strain
tensors, 
where a superscript $T$ denotes the transpose. 
We also introduce  the velocity gradient $\vec{D} \equiv
[\dot{\vec{F}} \vec{F}^{-1} + (\dot{\vec{F}} \vec{F}^{-1})^T]/2$,
where the superimposed dot represents the time derivative.
When a solid is \emph{incompressible}, every deformation is isochoric,
so that
\be \la{incomp}
J = 1, \qquad \text{tr }\vec{D} = 0.
\en
We recall that  the dissipated power per unit volume is 
$
\vec{\sigma}(t)  \vec{\cdot D}(t) \equiv \text{tr} \left(\vec{\sigma}(t)
\vec{D}(t)\right)
$,
so that $E_d$, the dissipated energy per unit volume over a period $T$,
is given by
\be
E_d = \dfrac{1}{T} \int_0^{T} 
 \vec{\sigma}(t)  \vec{\cdot D}(t) \text{d}t.
\en
Next we present the Abaqus FEA model.
 Section 4.8.2 of the Abaqus Theory Manual (Hibbit et al. 2007) gives
the following constitutive relation to model nonlinear viscoelastic
effects,
 \be \la{abaqus}
 \vec{\sigma}(t) = \vec{\sigma}_{e}(t) + 
    \text{SYM} \left\lbrace \vec{F}(t) \left[\int_{0}^{t}
\dfrac{J(s)}{J(t)} \dot{G}(t - s)
     \vec{F}^{-1}(s) \vec{\sigma}_e(s) \vec{F}(s) \text{d}s \right]
\vec{F}^{-1}(t)\right\rbrace, 
\en
where $\vec{\sigma}_e$ is the instantaneous Cauchy stress response
(elastic response 
at very short times) and  $G$ is the so-called memory kernel,
characterizing the stress relaxation 
(with $G(0)  = 1$).
Also, ``SYM''
denotes the symmetric part of the bracketed term; hence $\vec{D} =
\text{SYM} \{\dot{\vec{F}} \vec{F}^{-1}\}$.
This constitutive relation is valid for compressible as well as
incompressible solids, because in that 
latter case the hydrostatic term $-\hat{p}\vec{I}$ in $\vec{\sigma}_e$
(where $\hat{p}$ is a 
Lagrange multiplier) does not produce work, neither in the instantaneous
response, nor in the 
history term, as expected. 

For incompressible solids, $J=1$ at all times and $\vec{\sigma}_e$ has
the general form:
\be 
\vec{\sigma}_e = -\hat{p}\vec{I} + \psi_1 \vec{B} + \psi_2 \vec{B}^2,
\en
where $\psi_1$, $\psi_2$ are scalar functions of $t$ and of the first
and second principal strain
invariants.
Then \eqref{abaqus} reduces to 
\begin{multline} \la{abaqus_incomp}
 \vec{\sigma}(t) = - p(t) \vec{I}  + \psi_1(t) \vec{B}(t) + \psi_2(t)
\vec{B}(t)^2 \\
 + \sum_{i=1}^2   \text{SYM} \left\lbrace \vec{F}(t) \left[\int_{0}^{t}
\dot{G}(t - s)
   \psi_i(s)  \vec{C}(s)^{i} \text{d}s \right]
\vec{F}^{-1}(t)\right\rbrace, 
\end{multline}
where $p(t) = \hat{p}(t) + \int_0^t \dot{G}(t-s) \hat{p}(s) \text{d}s$ 
is arbitrary and remains to be determined from initial/boundary
conditions. 

The Abaqus FEA model is reminiscent of, and similar to a
well-established model of finite viscoelasticity, 
namely the Pipkin-Rogers model (Pipkin and Rogers 1968), which
reads in the
incompressible case
as (Wineman 1972, Johnson et al. 1996)
\be
\vec{\sigma}(t) =  -p(t)\vec{I} + \vec{F}(t) \left\{
\vec{R}[\vec{C}(t),0] + 
  \int_0^t \dfrac{\partial}{\partial(t-s)}(\vec{R}[\vec{C}(s),
t-s])\text{d}s \right\} \vec{F}(t)^T.
\la{qlve}
\en
Here $p$ is a Lagrange multiplier resulting from the internal constraint
of incompressibility and 
$\vec{R}$ is a strain dependent tensorial relaxation function, with the
general form
\be
\vec{R} = \phi_0 \vec{I} + \phi_1 \vec{C} + \phi_2 \vec{C}^2,
\en 
where $\phi_0$, $\phi_1$, $\phi_2$ are scalar functions of $t$ and of
the first and second principal strain
invariants.

By an appropriate choice of the $\phi_i$, the Pipkin-Rogers model
reduces to the so-called 
Quasi-Linear Viscoelastic (QLV) model, which has proved to be a most
successful 
phenomenological model for the behavior of non-linear viscoelastic
solids, see references in 
(Johnson et al. 1996, Wineman and Rajagopal 2008).
Equation (\ref{qlve}) is derived rigorously by successive 
approximations from the basic physical requirements 
governing the behavior of solids with memory 
(such as the principle of determinism and local action, 
the principle of material objectivity, etc.), 
see the review (Drapaca et al. 2007).

However we notice upon inspection of \eqref{abaqus_incomp} and
\eqref{qlve} that 
there are at least two differences between the models suggesting that
the Abaqus
FEA model does not rely on physical principles.
First, the integral term in equation \eqref{abaqus}
is generally non-symmetric, in contrast with the integral term in
equation \eqref{qlve}.
This is taken care of, in a somewhat arbitrary and \emph{ad hoc} manner,
by using the ``SYM'' operator.
Also, the history (time integral) term in the Abaqus FEA model
terminates with $\vec{F}(t)^{-1}$ in 
contrast with  the history term in the  the QLV model, which terminates
with $\vec{F}(t)^T$.

To emphasize the differences between each model, we henceforth focus on
a viscoelastic 
incompressible solid which has an instantaneous response modeled by the
neo-Hookean 
stress-strain relationship (already implemented into Abaqus), 
\be \vec{\sigma}_e = - p \vec{I} + \mu_0\vec{B},
\en
 where $\mu_0>0$ is a constant, the initial shear modulus. 
Also, the time relaxation of the solid is assumed to be governed by a
one-term 
Prony series expansion, so that after an infinite time the shear modulus
settles at the value 
$\mu_\infty$, say:
\be \la{prony}
G(t) = \dfrac{\mu_\infty}{\mu_0} + \left( 1 -
\dfrac{\mu_\infty}{\mu_0}\right)\ee^{-t/\tau}.
\en

 In that case, the Abaqus FEA model gives the identifications $\psi_1 =
\mu_0$, $\psi_2 = 0$, so that
\be \la{abaqus_neo}
 \vec{\sigma}(t) = - p(t) \vec{I}  + \mu_0 \vec{B}(t) 
 + (\mu_\infty - \mu_0 ) \text{SYM} \left\lbrace \vec{F}(t)
\left[\int_{0}^{t} \dfrac{\ee^{-(t - s)/\tau}}{\tau}
    \vec{C}(s)  \text{d}s \right] \vec{F}^{-1}(t)\right\rbrace.
\en

In contrast, the QLV model gives the identifications $\phi_0 =
\mu_0 G$, $\phi_1 = 0$, $\phi_2 = 0$, and the history term turns out
to be integrable, to give eventually
\be \la{qlv_neo}
 \vec{\sigma}(t) = - p(t) \vec{I}  
 + [\mu_\infty + (\mu_0 -
\mu_\infty) \ee^{-t/\tau}] \vec{B}(t).
\en
Here the viscoelastic effect is at its most limpid, whereas it seems
quite complicated in \eqref{abaqus_neo}.
We now look at two simple examples of motions.


\section{Equi-biaxial extension}


The equi-biaxial extension test of an incompressible solid is described
by the following motion,
\be
x_1 = \lambda(t) X_1,\qquad x_2 = \lambda(t)^{-1/2} X_2,\qquad x_3 =
\lambda(t)^{-1/2}  X_3,
\en
where $\lambda(t)$ is the stretch ratio in the direction of extension.
Then clearly the deformation gradient is diagonal,
\begin{equation}
\vec{F}(t) = \text{Diag} \left[ \lambda (t), \lambda (t)^{-1/2},
\lambda(t)^{-1/2} \right].
\label{DeformationGradient}
\end{equation}

Assuming now that the solid is subject to \emph{uni-axial
tension}: $\sigma_{[11]} \ne 0$, $\sigma_{[22]} = \sigma_{[33]} = 0$,
allows us to compute the Lagrange multiplier $p$.
Then we find the resulting Cauchy stress non-zero component is
\be
\sigma_{[11]}(t) = \mu_0 \left[\lambda ^2(t) - \lambda(t)^{-1} \right] +
  (\mu_\infty - \mu_0 ) \int_{0}^{t}  \dfrac{\ee^{-(t - s)/\tau}}{\tau}
\left[\lambda ^2(s) - \lambda(s)^{-1}\right]  \text{d}s,
\en
for the Abaqus FEA model, and
\be
\sigma_{[11]}(t) = [\mu_\infty + (\mu_0 -
\mu_\infty) \ee^{-t/\tau}]
    \left[\lambda ^2(t) - \lambda(t)^{-1}\right],
\en
for the QLV model.
The difference between each model is now clearly apparent, and it  
reflects on the dissipated power, as we know confirm numerically.

From an experimental point of view, it is  common practice to have a
dynamic displacement 
superimposed on a large static deformation. 
Here we consider that the neo-Hookean solid is strained in equi-biaxial
tension by 30\%
from time $t=0$ to time $t=1$, 
and then made to oscillate with an amplitude of 20\%:
\be
\lambda(t) = 
\begin{cases}
1 + 0.3 t, \qquad 0 \leq t \leq 1, \\
1.3 + 0.2 \sin \omega (t - 1), \qquad t > 1.
\end{cases}
\la{LambdaSin}
\en
\begin{figure}
\centering
\includegraphics[width=0.8\textwidth, height=0.4\textwidth]{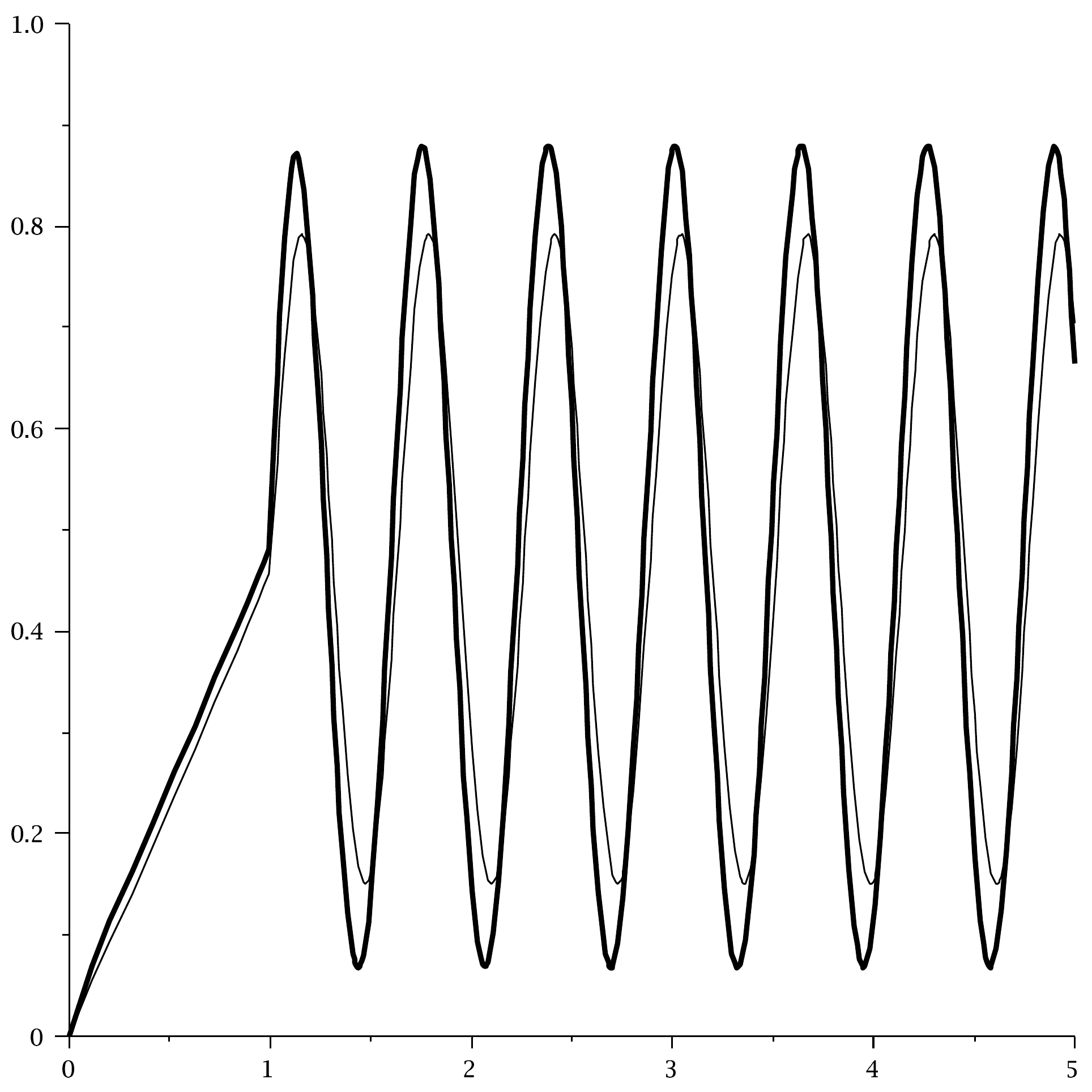}
\caption{Variations of the axial tension $\sigma_{[11]}$ with time,
for the Abaqus FEA model (thick curve) and for the QLV model
(thin curve) in the case of equi-biaxial tension.}
\label{fig:Sigma11}
\end{figure}
For the remaining parameters we pick
\begin{equation} \la{parameters}
\mu_\infty / \mu_0 = 0.5, \qquad \tau = 0.1 \text{ s}, \qquad \omega =
10.0 \text{ s}^{-1},
\end{equation}
or we keep $\omega$ as a free parameter when we investigate the
frequency 
dependence of the dissipated energy.

Hence Figure \ref{fig:Sigma11} shows that at those 
values, the predictions from the Abaqus model differ widely from those
of the QLV model.

\begin{figure}
\centering
\includegraphics[width=0.8\textwidth]{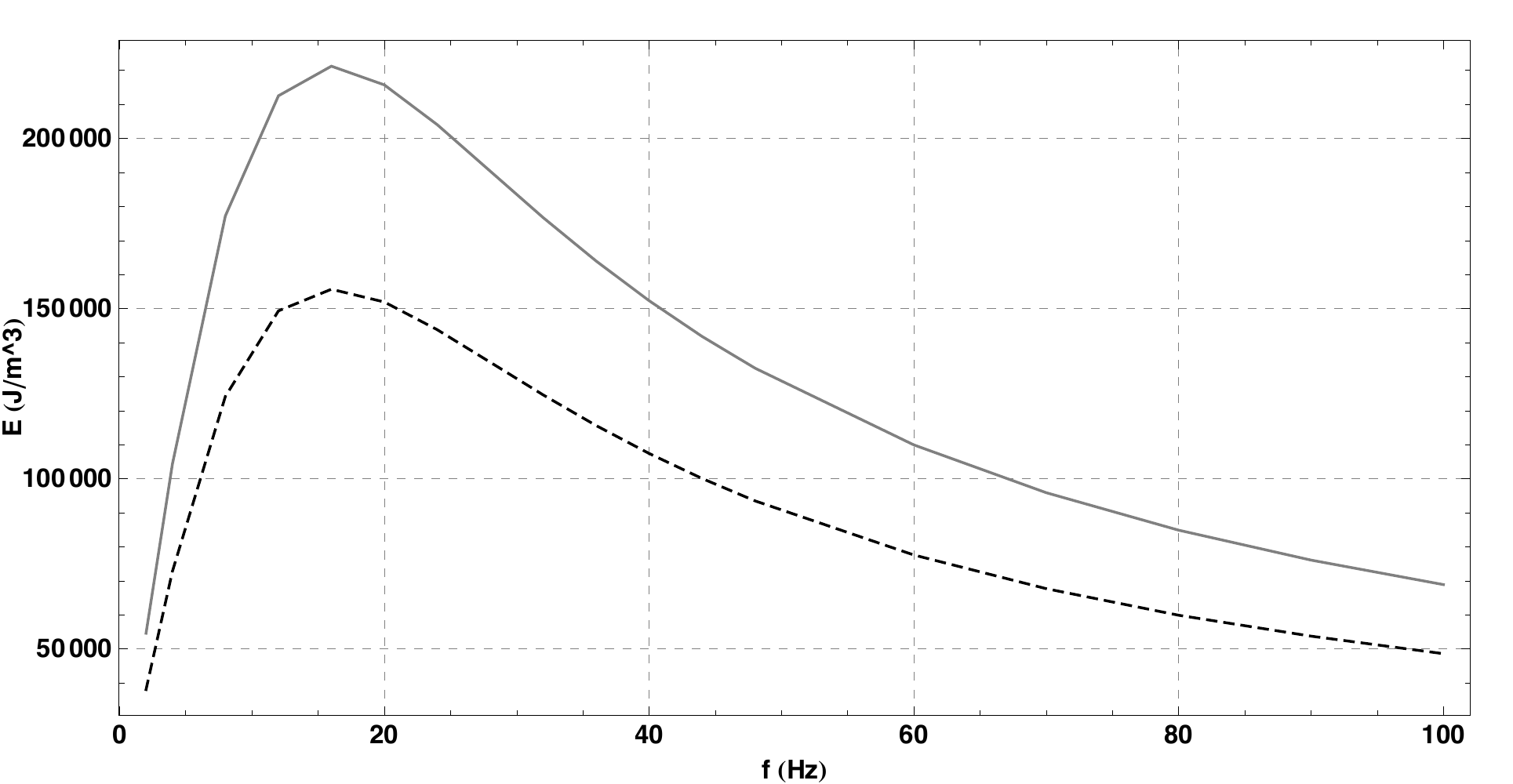}
\caption{Energy dissipation per unit volume over a single period for the
Abaqus model 
(gray, solid curve) and for the QLV model (black, dashed curve) in the
case of equi-biaxial 
tension.}
\label{fig:TensionDissipatedEnergy}
\end{figure}

The numerical simulations are performed with Abaqus 6.7-1 on a
bidimensional square element of 1 m side; 
we use a single CPS4 element (4 nodes, bilinear, plane stress) with an
implicit solution scheme.
Figure \ref{fig:TensionDissipatedEnergy} shows clearly that the Abaqus
model overestimates 
the energy dissipation compared to the QLV model.
%


\section{Simple Shear}


We now consider a simple shear of amount $\gamma$ say, 
\be
x_1 = X_1 + \gamma(t) X_2, \qquad x_2 = X_2, \qquad x_3 = X_3,
\en
for which the deformation gradient is not symmetric,
\begin{equation}
\vec{F}(t)=%
\begin{bmatrix}
1 & \gamma(t) & 0 \\ 
0 & 1 & 0  \\ 
0 & 0 & 1%
\end{bmatrix}.
\label{SimpleShear}
\end{equation}

Simple calculations reveal that a sheared solid described by the Abaqus
FEA model
or by the QLV model is in a state of \emph{plane stress} 
($\sigma_{[ij]} \ne 0$ for $i,j = 1,2$; $\sigma_{[3j]} = 0$ for
$j=1,2,3$) when the Lagrange 
multiplier $p$ is taken as
\be
p(t) = \mu_\infty - (\mu_\infty - \mu_0)\ee^{-t/\tau}.
\en
Then we find for the Abaqus FEA model, 
\begin{align} \la{shear_abaqus}
& \sigma_{[11]}(t) = 
 (\mu_\infty - \mu_0) \int_0^t
 \dfrac{\ee^{-(t-s)/\tau}}{\tau}
   \gamma(s) \gamma(t) \text { d} s,
  \notag \\
& \sigma_{[12]}(t) = \mu_0 \gamma(t) 
 +  (\mu_\infty - \mu_0) \int_0^t
 \dfrac{\ee^{-(t-s)/\tau}}{2\tau}
  \gamma(s)\left[2 +
\gamma(s)\gamma(t) - \gamma(t)^2\right]\text{d}s, 
   \notag \\
& \sigma_{[22]}(t) = 
  (\mu_\infty - \mu_0) \int_0^t
 \dfrac{\ee^{-(t-s)/\tau}}{\tau}
   \gamma(s)\left[\gamma(s)-\gamma(t)\right
]\text{d}s,
\end{align}
and for the QLV model
\begin{align} \la{shear_qlve}
& \sigma_{[11]}(t) = 
[\mu_\infty - (\mu_\infty -
\mu_0)\ee^{-t/\tau}]\gamma(t)^2, 
\notag  \\
& \sigma_{[12]}(t) = [\mu_\infty - (\mu_\infty -
\mu_0)\ee^{-t/\tau}]\gamma(t), 
\notag  \\
& \sigma_{[22]}(t) = 0.
\end{align}
These latter expressions are intuitively expected for the neo-Hookean
solid,  because for its instantaneous response at very short times, 
the Cauchy stress has
components $\sigma_{e[11]} = \mu_0 \gamma^2$,  
$\sigma_{e[12]} = \mu_0 \gamma$,
$\sigma_{e[22]} = 0$.
In contrast, the Abaqus FEA model gives rise to a $\sigma_{[22]}$
component purely due to viscoelastic effects.
This behavior is clearly non-physical, for in the case of a quasi-static
loading, no traction $\sigma_{[22]}$ 
is required to shear a neo-Hookean solid, 
and yet the relaxation process would create such a component \emph{ex
nihilo}!
\begin{figure}
\centering
\includegraphics[width=0.8\textwidth, height=0.4\textwidth]{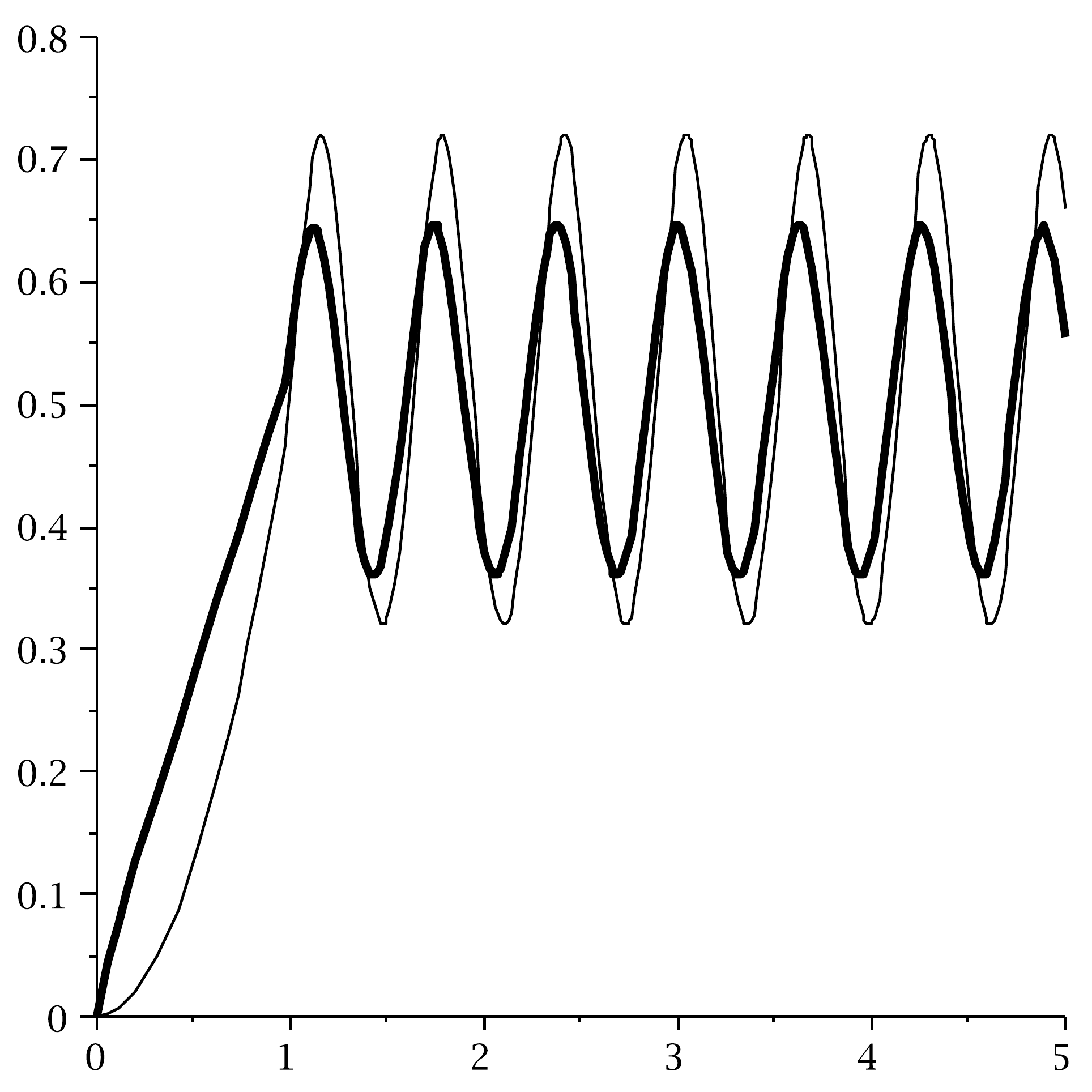}
\caption{Variations of $\sigma_{[12]}(t)/\mu_0$
for the Abaqus analytical model (thick curve) 
and the QLV model (thin
curve).}
\label{fig:s12}
\end{figure}
\begin{figure}
\centering
\includegraphics[width=0.8\textwidth, height=0.4\textwidth]{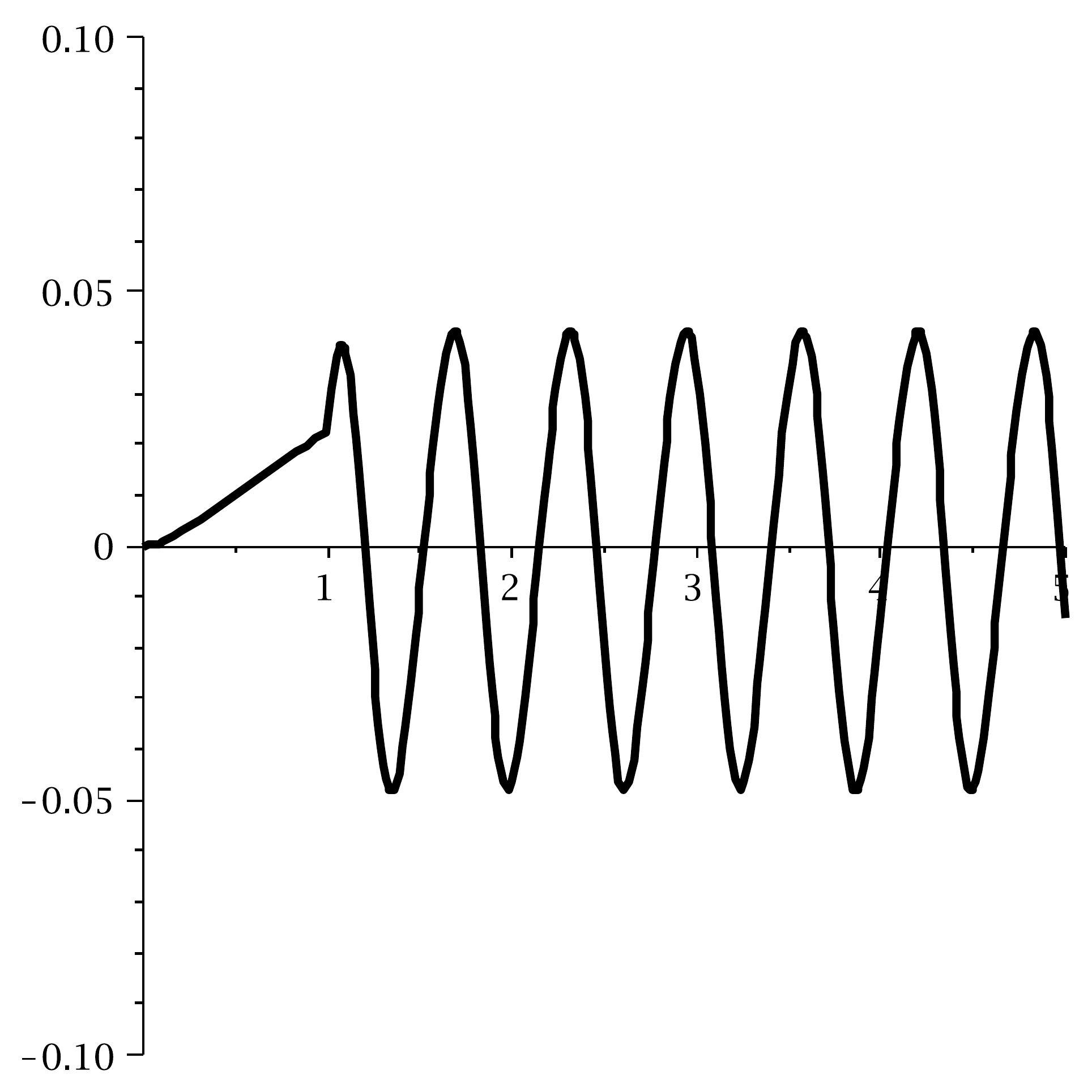}
\caption{Variations of $\sigma_{[22]}(t)/\mu_0$ for the
Abaqus FEA model (thick curve) and the QLV model (thin curve).
The latter does not show clearly on the graph, because it is equal to
zero at all times.}
\label{fig:s22}
\end{figure}
These discrepancies  have grave consequences for the dissipated power
because here
\begin{equation} \la{shear_D}
\vec{D}(t)=%
\begin{bmatrix}
0 & \dot{\gamma}(t)/2 & 0 \\ 
\dot{\gamma}(t)/2 & 0 & 0  \\ 
0 & 0 & 0
\end{bmatrix},
\quad \text{so that} \quad 
\vec{\sigma \cdot D} = \dot{\gamma}(t) \sigma_{[12]}(t),
\end{equation}
which is obviously not the same depending on the model considered.

\begin{figure}
\centering
\includegraphics[width=0.8\textwidth]{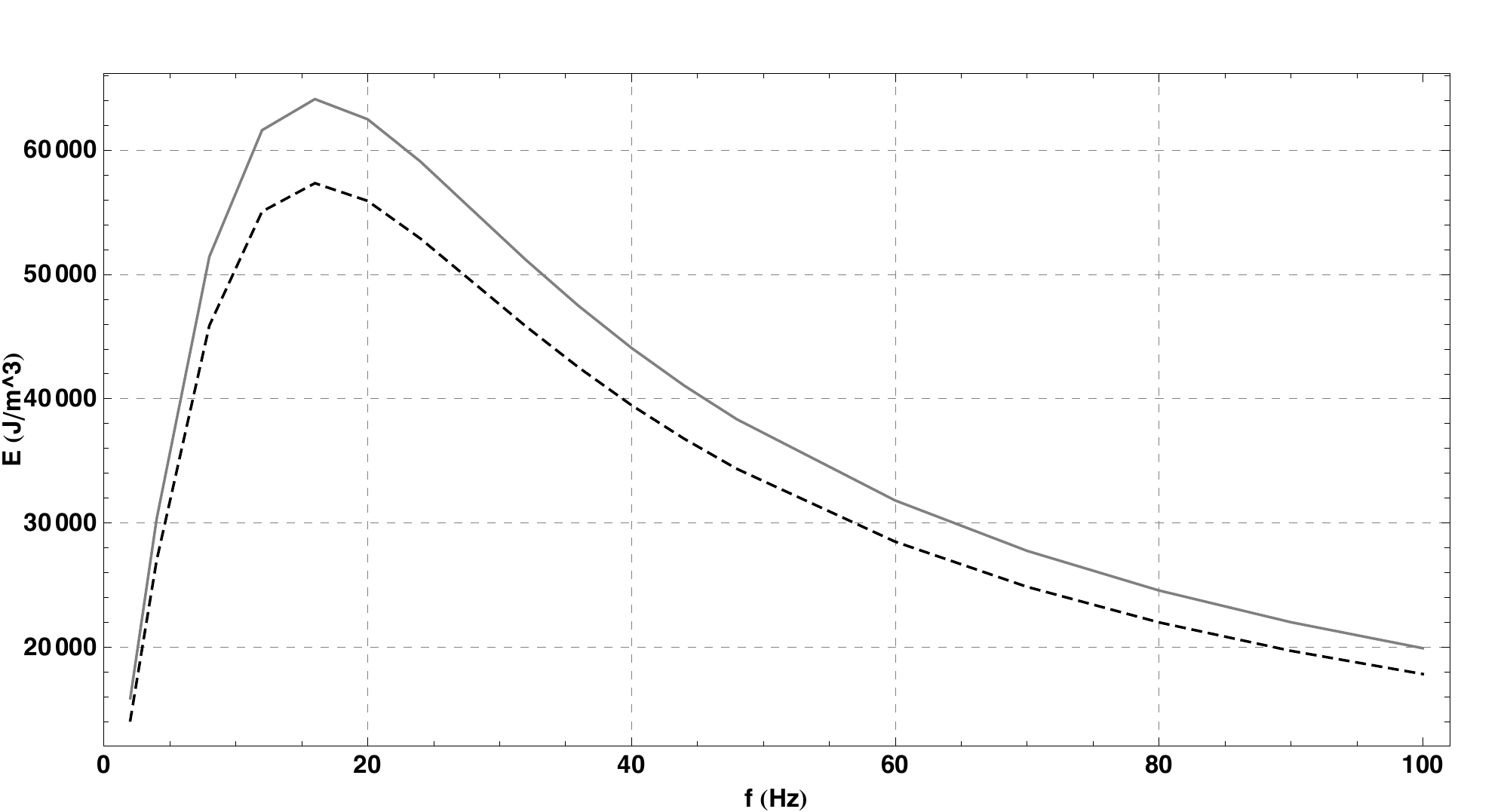}
\caption{Energy dissipation per unit volume over a single period in the
case of 
the Abaqus model (gray, continuous curve) and of the  QLVE model (black,
dashed curve) for 
simple shear.}
\label{fig:ShearDissipatedEnergy}
\end{figure}

We confirm these findings by testing the Abaqus FEA software against the
current formulas 
\eqref{shear_abaqus}-\eqref{shear_D}.
We take the amount of shear $\gamma(t)$ to vary as
\be
\gamma(t) = 
\begin{cases}
t, \qquad 0 \leq t \leq 1 \\
1 + 0.2 \sin\left[ \omega  \left(t-1\right)\right], \qquad t > 1,
\end{cases}
\la{GammaSin}
\en
with the other parameters given by \eqref{parameters}.
Figures \ref{fig:s12} and \ref{fig:s22} display the
variations of the $\sigma_{[12]}$ and 
$\sigma_{[12]}$ components computed from \eqref{shear_abaqus} and
\eqref{shear_qlve}.
Finally, we find that the Abaqus model overestimates the energy
dissipation with respect to
the QLV model, see Figure \ref{fig:ShearDissipatedEnergy}.
Using the commercial software simulation, we recovered the thick
curves, which confirms our opinion that equation
\eqref{abaqus} is actually implemented in the commercial code..

%
%
%
%
%
%


\section*{Acknowledgements}


We thank the Universit\`a
di Roma ``La Sapienza'' (Italy), the CNRS (France), and the Royal
Society (UK) for their support.



\end{document}